\documentclass[letterpaper, 10 pt, conference]{ieeeconf}

\IEEEoverridecommandlockouts                             

\usepackage{graphicx}
\usepackage{tabularx}
\usepackage{flushend}
\usepackage{xcolor}
\usepackage{booktabs}
\usepackage{array} 
\usepackage{comment}
\usepackage{amssymb}
\usepackage{amsmath}

\newcommand{\etal}{et al.}

\title{\LARGE \bf
Clustering Social Touch Gestures for Human-Robot Interaction
}

\author{Ramzi Abou Chahine$^{1}$, Steven Vasquez$^{2}$, Pooyan Fazli$^{3}$, and Hasti Seifi$^{4}$%
\thanks{$^{1}$Ramzi Abou Chahine is with the School of Computing Science, 
University of East Anglia, Norwich, Norfolk NR4 7TJ, UK
        {\tt\small r.abou-chahine@uea.ac.uk}}%
\thanks{$^{2}$Steven Vasquez is with the Department of Computer Science, San Francisco State University,
        San Francisco, CA 94132, USA
        {\tt\small svasquez7@mail.sfsu.edu}}%
\thanks{$^{3}$Pooyan Fazli is with the School of Arts, Media and Engineering at Arizona State University, Tempe, AZ 85281, USA
        {\tt\small pooyan@asu.edu}}
\thanks{$^{4}$Hasti Seifi is with the School of Computing and Augmented Intelligence at Arizona State University, Tempe, AZ 85281, USA
        {\tt\small hasti.seifi@asu.edu}}
}

\begin{document}

\maketitle
\thispagestyle{empty}
\pagestyle{empty}

\begin{abstract}
 Social touch provides a rich non-verbal communication channel between humans and robots. 
 Prior work has identified a set of touch gestures for human-robot interaction and  described them with natural language labels (e.g., stroking, patting). Yet, no data exists on the semantic relationships between the touch gestures in users' minds. To endow robots with touch intelligence, we investigated how people perceive the similarities of social touch labels from the literature. In an online study, 45 participants grouped 36 social touch labels based on their perceived similarities and annotated their groupings with descriptive names. We derived quantitative similarities of the gestures from these groupings and analyzed the similarities using hierarchical clustering. The analysis resulted in 9 clusters of touch gestures formed around the social, emotional, and contact characteristics of the gestures. We discuss the implications of our results for designing and evaluating touch sensing and interactions with social robots. 
\end{abstract}

\section{Introduction}
Social touch has been an active area of research for human-robot interactions (HRI) in the last decade.
Social touch gestures refer to different ways that people use touch to communicate information or emotion and bond with other humans or robots~\cite{huisman2017social}. 
For example, one may tap a robot's arm to get its attention or hug a robotic pet when stressed. A companion robot may stroke a user's hand to convey emotional support or guide the user's action by pushing their hand. 
Previous work has derived a set of social touch gestures and their definitions based on user interactions with robotic pets~\cite{yohanan2012role}. Others designed and evaluated touch interactions with humanoid robots~\cite{fitter2014analyzing,burns2022touchperception}. The touch gestures from these studies have guided the development and evaluation of touch sensors for robots, helped examine user experience of robot-initiated touch, and informed the design of robot response to user touch.

\begin{figure}[t]
    \centering
    \includegraphics[
        width = \linewidth
    ]{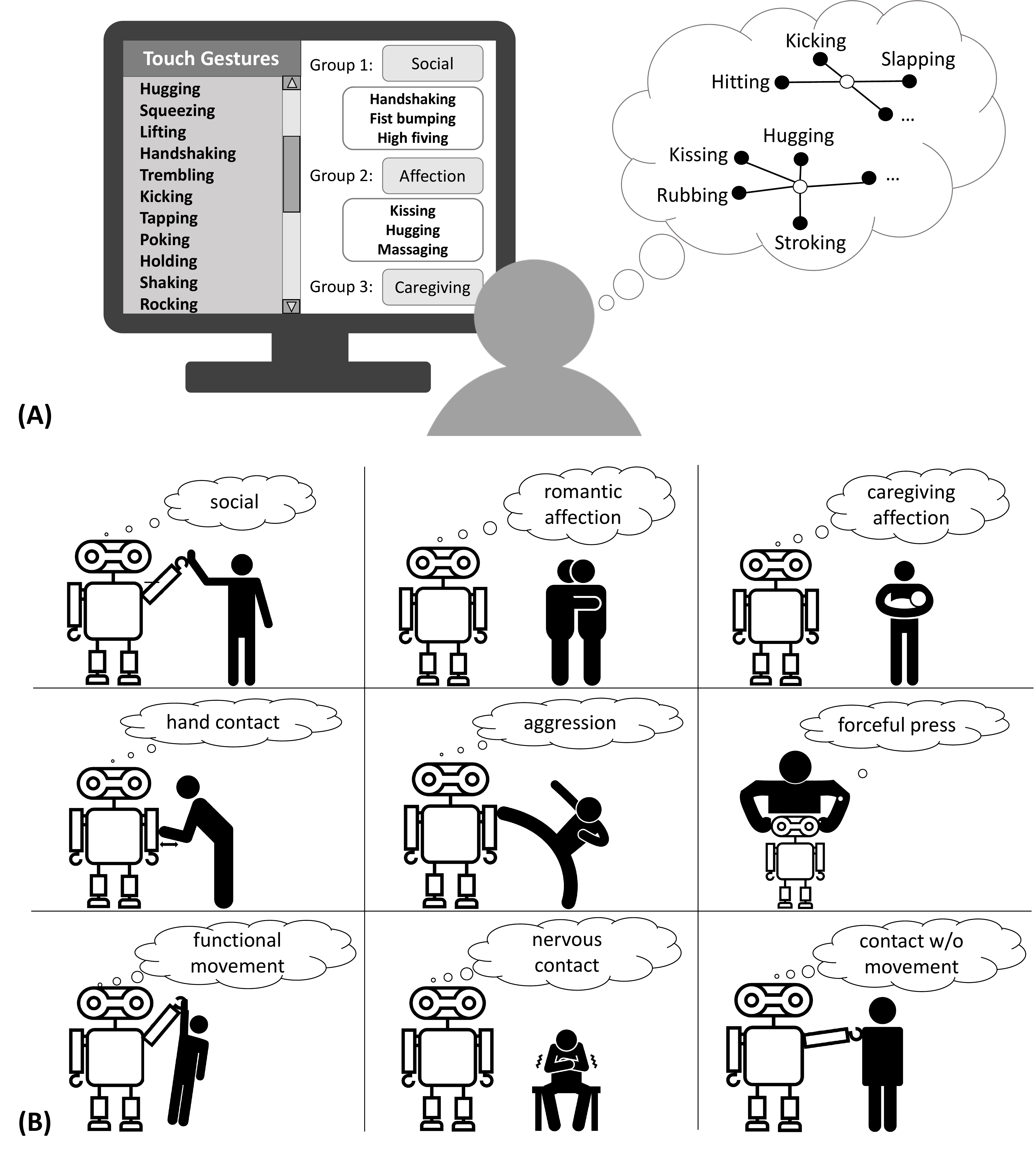}

    \label{fig:teaser}
       \caption{In our study, users grouped social touch labels based on their perceived similarities (A). The resulting touch clusters can be used by robots to interpret and perform touch interactions with people (B).}
\end{figure}

 Despite the abundance of interest in social touch communication, the semantic relationship(s) between various touch gestures remains unclear. Some gestures may be very similar or even identical in their contact characteristics (e.g., \textit{tapping} vs.\ \textit{patting}), while others may be similar considering the intended emotion or social context. People develop a mental structure for the semantics of touch gestures and their relationships. This mental structure shapes people's perception, interpretation, and use of touch~\cite{hertenstein2009}. 
Charting the relationship between social touch gestures can help HRI researchers select touch gestures for their studies (e.g., touch sensor evaluation) and develop robots that use touch in a socially intelligent manner. Yet, little data exists in the literature about how people perceive the relationships between social touch gestures.

As a first step toward addressing this gap, we asked how people perceive similarities of social touch labels (e.g., \textit{stroking}, \textit{hugging}). Touch is inherently multimodal and people can have unique styles in applying a touch gesture~\cite{jung2014touching}. On the other hand, people often use natural language labels to refer to archetypal features of a touch gesture. The touch labels are also used in HRI studies to ask users to contact a robot (or a sensor) in a certain way~\cite{jung2014touching,burns2022touchperception} or to analyze user interactions with a robot~\cite{yohanan2012role}. The study of natural language labels for emotions has helped capture users' cognitive structure, leading to a circumplex model for affect~\cite{russel1980}. Thus, as a first step, we investigated the semantic structure of social touch labels in the users' minds in this paper. 

To chart the relationship between touch labels, we 
ran an online card sorting study with 56 users over Amazon Mechanical Turk (Figure~\ref{fig:teaser}-A). 
The participant received the labels and definitions for 36 touch gestures from the literature, sorted them into 4, 8, and 12 groups successively based on their similarities, and provided descriptive names for each group. From this data, we identified 11 outliers by manually reviewing the data as well as analyzing the responses quantitatively. 
Then, we created a dissimilarity matrix for the 36 touch gestures with the data of the remaining 45 participants and applied agglomerative hierarchical clustering on the dissimilarity matrix.  
Furthermore, we analyzed the descriptive names that the participants had for their groupings using open codes (e.g., \textit{social}, \textit{aggressive}) and calculated the frequency of the codes for the gestures.

Based on the above analysis, we contribute 9 clusters for social touch gestures and the distribution of the top codes for each cluster. Using this data, we interpret the 9 clusters to capture the types of touch as follows: 
(1) social, (2) romantic affection, (3) caregiving affection, (4) hand contact, (5) aggression, (6) forceful press, (7) functional movement, (8) nervous contact, and (9) contact without movement (Figure~\ref{fig:teaser}-B). 
Our results suggest that people primarily group touch gestures based on their social, emotional, and contact characteristics. 
These results provide the first data on cognitive structure(s) that people use in interpreting and conceptualizing social touch. 
We discuss how the results can help design and evaluate a robot to sense, interpret, and communicate via touch.

\section{Related Work}
\subsection{Social Touch in HRI}
The literature on social touch ranges from communication between humans to interactions between humans and robots. 
Hertenstein \etal~studied how dyads use social touch gestures to communicate different emotions and found that people can decode the intended emotions with great accuracy when being touched~\cite{hertenstein2009}. Similar studies of human-human touch suggest that touchers can subtly but significantly vary contact attributes of their touch actions to communicate distinct messages~\cite{xu2021subtle}. HRI researchers have replicated Hertenstein \etal's work to investigate how users and robots can use touch to communicate emotions. Some studies examined how humans communicate emotions to robots~\cite{yohanan2012role,cang2015different,andreasson2018affective}, while others examined whether a robot can communicate emotions to humans via touch~\cite{willemse2019social,teyssier2020conveying,seifi2023firsthand}. 

Social touch gestures have also informed the development and evaluation of tactile skins for robots.
Previous work in this area has proposed touch sensors with a novel working principle~\cite{Choi2022Learning}, sensors resembling the feel and structure of human skin~\cite{teyssier2021human}, and low-cost do-it-yourself sensors for specific applications such as companion robots for children with autism~\cite{burns2022touchperception}.
To evaluate the sensor's efficacy, researchers select a set of social touch gestures and ask users to touch the sensor accordingly. Data from user contact with the sensor is then used to classify the gestures. 

A variety of touch gestures are reported in the above studies. Yohanan and MacLean proposed a touch dictionary with labels and definitions for 30 touch gestures based on videos of user interactions with a furry lap-sized robot~\cite{yohanan2012role}.
This dictionary has been widely used in social touch studies~\cite{burns2022touchperception, cang2015different, jung2017automatic}. Others mentioned additional touch gestures that are relevant to interactions with humanoid robots such as \textit{fist bumping}~\cite{pelikan2020you,rognon2022online,prasad2022mild},
\textit{handshaking}~\cite{wang2011handshake,ammi2015haptic,rognon2022online}, or \textit{kicking}~\cite{salter2007using,hertenstein2009,li2016human}.
To inform future work in this area, we collected common touch gestures from prior studies and examined how people conceptualize the relationship between these gestures.

\subsection{Identifying Perceptual and Semantic Clusters}
The psychophysics and interaction design literature has developed methods for estimating perceptual and semantic similarities of items through user studies.
The pairwise rating method asks participants to rate the similarity of pairs of items in the set~\cite{eurohaptics22, park2011perceptual}.
This method is effective for a small set of items (e.g., $<15$) but it is prone to noise from local judgments and does not scale to large item sets~\cite{tsogo2000multidimensional,spam2013versatility}.
The sorting methods, known as card sorting or cluster sorting, ask participants to group items into clusters based on their similarities. This process can be repeated with an increasing number of groups to obtain a fine-grained similarity matrix~\cite{ternes2008designing,rugg1997sorting}. This method allows for collecting cognitive similarities of large item sets~\cite{russel1980,seifi2020capturing}. 
The similarity matrix is further analyzed using dimensionality reduction~\cite{russel1980, eurohaptics22, park2011perceptual} or clustering techniques~\cite{mun2019perceptual,seifi2020capturing}. Following this methodology, we used iterative cluster sorting and asked users to name their groups to obtain semantic clusters for social touch labels.

Natural language labels have been used to capture lay users' cognitive structure for sensory and emotional items. The circumplex model of affect by Russell~\cite{russel1980} is based on a series of studies that use natural language labels for emotions. 
Also, studies of social touch often rely on user understanding of natural language labels for touch. In these studies, users receive labels for a set of social touch gestures (e.g., \textit{tapping}, \textit{stroking}) and are asked to touch the robot accordingly~\cite{jung2017automatic,burns2022touchperception}. 
Similarly, studies on human-human and human-robot emotional communication sometimes provide a list of touch gesture labels for users to choose from, before applying the gestures~\cite{hertenstein2009,xu2021subtle}. These studies may provide short definitions for each touch gesture e.g., from the touch dictionary by Yohanan and MacLean~\cite{yohanan2012role}. These studies rely on the users' knowledge of natural language labels for touch gestures. We follow a similar approach in our work to capture user's cognitive structure and similarities of social touch gestures.

\begin{table*}[ht]
\centering
\footnotesize
\caption{The 36 gestures that we used in the online study; 29 gestures are from the touch dictionary by Yohanan and MacLean~\cite{yohanan2012role}, and we added 7 touch gestures from the social touch literature. The newly added gestures are marked with a *. Entries are listed alphabetically by \textit{Gesture Label}.} 
\label{tab:touchdic}
\begin{tabular}{p{2cm}p{6cm}p{2cm}p{6cm}}
\toprule
Gesture Label    & Gesture Definition & Gesture Label & Gesture Definition\\
\cmidrule(r){1-2}
\cmidrule(r){3-4}
Contacting Without Movement & Any undefined form of contact with something that has no movement. & Pinch   & Tightly and sharply grip something between your fingers and thumb. \\
Cradle  & Hold something gently and protectively.  & Poke  &  Jab or prod something with your finger.   \\
Finger Interlocking$^{*}$          & Interlace fingers of one hand
with another hand with palms.        & Press  &  Exert a steady force on something with
your flattened fingers or hand.     \\
Fist Bumping$^{*}$ & Lightly tap clenched fists together.  & Pull   & Exert force on something by taking hold of it in order to move it towards yourself.    \\
Grab & Grasp or seize something suddenly and roughly.  & Push   & Exert force on something with your hand
in order to move it away from yourself. \\
Handshaking$^{*}$ & Shake clasped hands. & Rock  & Move something gently back and forth or from side to side.    \\
High Fiving$^{*}$ & Slap upraised hands against each other. & Rub & Move your hand repeatedly back and forth on something with firm pressure.  \\
Hit &  Deliver a forcible blow to something.  & Scratch   &  Rub something with your fingernails. \\
Hold &  Grasp, carry, or support something with your arms or hands.  & Shake   & Move something up and down or side to side with rapid, forceful, jerky movements.\\

Hug & Squeeze something tightly in your arms. & Slap   & Quickly and sharply strike something
with your open hand. \\
Kicking$^{*}$ & Strike forcibly with a foot.    & Squeeze    &  Firmly press something between your
fingers or both hands. \\
Kiss   &  Touch something with your lip.   & Squish$^{*}$ & Press or beat into a pulp or a flat mass.  \\
Lift  &  Raise something to a higher position or level.   & Stroke    & Move your hand with gentle pressure over something, often repeatedly.  \\
Massage  & Rub or knead something with your hands.   & Swing    & Move something back and forth or from
side to side while suspended.  \\
Nuzzle  &  Gently rub or push against something with your nose or mouth.    & Tap    & Strike something with a quick light blow or blows using one or more fingers.\\
Pat  &   Gently and quickly touch something.   & Tickle    &  Touch something with light finger movements.\\
Pick & Repeatedly pull at something with one or more of your fingers. & Toss   & Throw something lightly, easily, or casually.    \\
Picking Up$^{*}$ & Take hold of and lift or move something. & Tremble   &  Shake against something with a slight rapid motion. \\

\bottomrule
\end{tabular}
\end{table*}

\section{Methods}
To study how people perceive similarities of social touch gestures, we compiled a list of touch gestures from the literature, designed an online questionnaire for grouping the touch gestures, and ran a data collection study on MTurk. 
\subsection{Touch gestures}
We compiled 36 social touch gestures that are used for interacting with humans or robots (Table~\ref{tab:touchdic}). 
We focused our scope on gestures that are used in at least two publications in the social touch and HRI literature. 
Specifically, we included 29 touch gestures from the touch dictionary by Yohanan and MacLean~\cite{yohanan2012role}.
Different subsets of these gestures are used in several other studies~\cite{silvera2012interpretation,jung2014touching,cang2015different}. 
We removed \textit{finger idly} from the touch dictionary as this gesture is not used in any other publication. 
We added seven other touch gestures that appeared in at least two publications including 
\textit{finger interlocking}~\cite{hertenstein2009,hertenstein2011gender}, 
\textit{fist bumping}~\cite{pelikan2020you,rognon2022online,prasad2022mild}
\textit{handshaking}~\cite{wang2011handshake,ammi2015haptic,rognon2022online}, 
\textit{high fiving}~\cite{cramer2009effects,hertenstein2009,fitter2014analyzing}, \textit{squishing}~\cite{burns2021getting}, \textit{kicking}~\cite{salter2007using,hertenstein2009,li2016human}, and \textit{picking up}~\cite{salter2006learning,salter2007using,burns2021getting}. 

We adapted the definitions provided in Yohanan and MacLean's touch dictionary by replacing the phrases related to their robotic pet (i.e., the Haptic Creature, or fur of Haptic Creature) with ``something'' in the definition. 
For example, we defined \textit{lifting} as ``raise something to a higher position or level.'' For the 7 actions that were not in the original touch dictionary, we created a definition with inspiration from sources such as the Britannica Encyclopedia. 

\subsection{Questionnaire}
\begin{figure}[h]
    \centering
    \includegraphics[width=\linewidth]{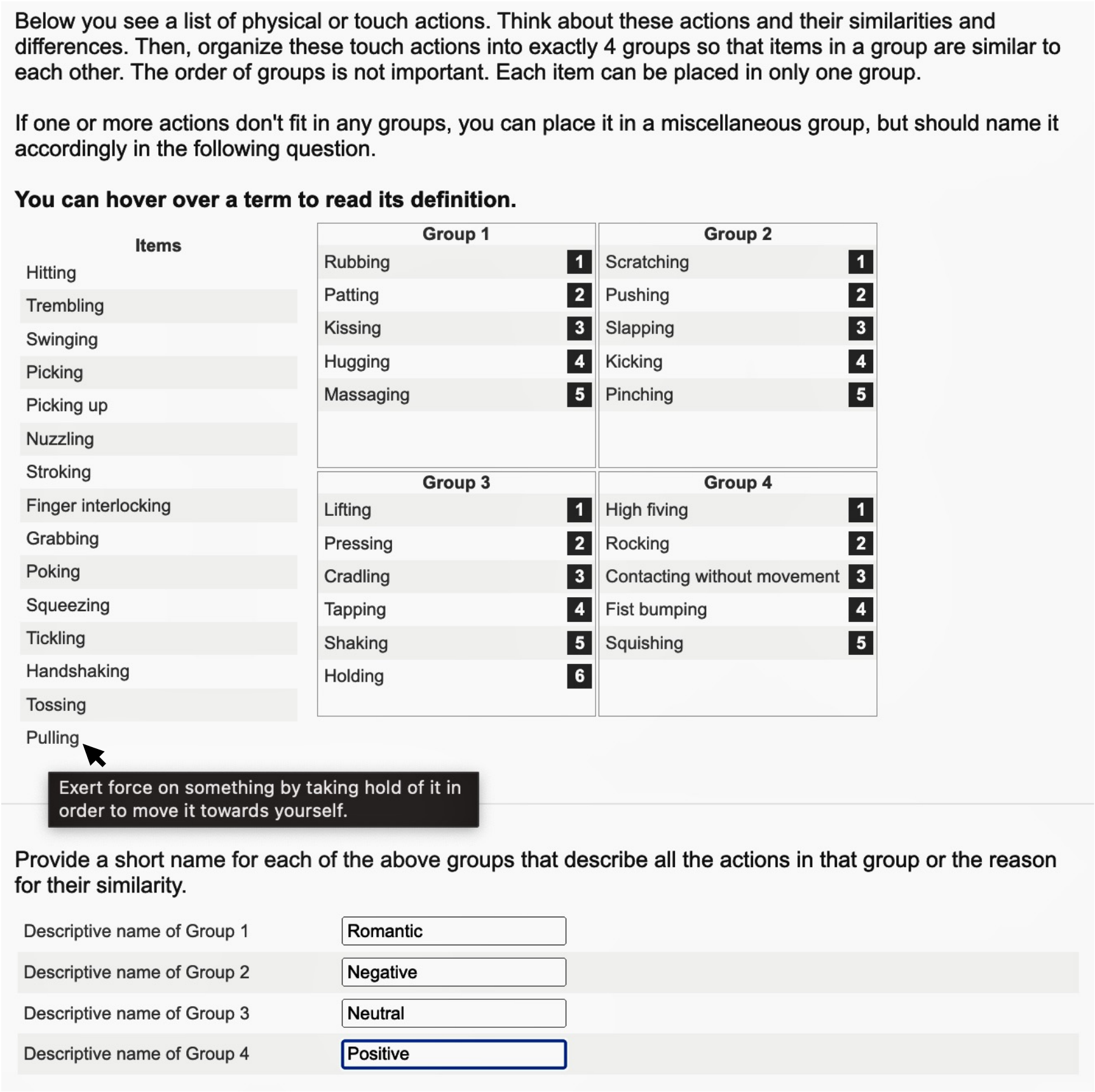}
    \caption{A screenshot of the questionnaire for grouping the touch gestures in our study. The image shows touch gestures that are divided into four groups, the remaining list of gestures for grouping, and example descriptive names from one of the participants.}
    \label{fig:survey}
\end{figure}

We designed a Qualtrics survey to collect user demographics and data on the similarity of touch actions (Figure~\ref{fig:survey}). 
The first page of the survey asked users to enter their demographic information including their age, gender, and country where they grew up.
The next three pages asked the users to divide the touch gestures into 4, 8, and 12 groups respectively. We call these 4 groupings, 8 groupings, and 12 groupings in the remainder of the paper. Each page showed the list of touch gesture labels in random order. The users could hover over a gesture's label to see its definition. The users were asked to group the touch gestures based on their likeness or similarity and provide a descriptive name for each group. Reasons for likeness were up to user interpretation. 
Having the users describe their groupings served multiple purposes. First, they helped us identify users' reasoning for the similarity of touch gestures. Second, the descriptive names served as an attention test and allowed us to detect those who did not do the task properly, e.g., if they organized the 36 touch gestures into random groups.

We devised the above procedure based on common practices in studies of similarity perception and social touch gestures in the literature.
First, the iterative cluster sorting method allowed us to collect user's holistic comparisons of the similarities of all 36 gestures. 
Second, following prior work on touch sensing and communication, the touch labels helped us abstract from a variety of styles that people use to apply the touch gestures (e.g., tapping one time or multiple times) to capture users' cognitive structure of the gestures. 

\section{Analysis and Results}
We collected a total of 88 responses through MTurk. 
Eligible turkers were required to have at least 5,000 completed tasks with a success rate of 97\% or higher and to speak English at the B2 level or higher. 
The participants were from the United States ($n=47$), followed by India (33), Brazil (7), and Japan (1). They self-identified as man ($n=60$), woman (26), or nonbinary (0). \noindent We analyzed their data in the following steps:
\vspace{0.2cm}

\begin{itemize}
\item\textbf{Identifying outliers.} We identified participants who did not follow the study instructions or appeared to group the touch gestures randomly (Section~\ref{sec:outliers}) and removed their data from the subsequent analysis. We also examined the effect of the participant's country of origin in their groupings.
\item\textbf{Coding descriptive names for the groups.} To identify the themes behind the user groupings, we coded the descriptive group names from the participants. This step resulted in 25 codes (e.g., social, aggressive) to capture user logic for their groupings (Section~\ref{sec:coding}). 
\item\textbf{Clustering touch gestures.} We calculated a dissimilarity matrix for the touch gestures based on the participants' groupings of the gestures. We then identified semantic groups by applying hierarchical clustering on the dissimilarity matrix. 
\item\textbf{Interpreting the clusters.} Finally, we counted how many times a code from Step II was applied to the touch gestures in each cluster. The results helped us interpret and label each of the 9 social touch clusters (Section~\ref{sec:clustering}).
\end{itemize}

Below we detail these steps and their results.

\subsection{Identifying outliers}
\label{sec:outliers}

We marked and removed outliers who did not follow the study instructions or their groupings and descriptive names appeared random in three steps.

In the first step, we removed data from 32 (out of 88) participants who did not follow the study instructions to provide descriptive names for each group. These responses either included no label, gibberish, or simply repeated the name of a term in each group. Also, the groupings did not seem to follow any logic and appeared random to the authors.

In the second step, one of the authors carefully examined all the responses from the remaining 56 participants and marked potential outliers for further analysis. 
The author marked cases where the description of group labels did not match with its gesture items. For example, if a participant grouped \textit{kissing}, \textit{nuzzling}, and \textit{stroking}  with \textit{hitting} and labeled them as ``fighting'', we marked this as an unusual group. By the end of this step. 16 participants with several unusual groupings were marked as potential outliers. 

In the third step, we calculated the similarity between the participants. We first created three similarity matrices based on the 4, 8, and 12 groupings. Each cell in these matrices showed similarity of the groupings provided by two participants ($56 \times 56$ matrix) in their 4, 8, or 12 groupings. To obtain the best matching between groups from two different participants, we calculated the Jaccard Index values for all pairs of groups provided by them (e.g., 8 pairs for the 4 groupings) and averaged the highest Jaccard values 
as a measure of the similarity of the two participants. Finally, we averaged the participant similarity values over the three matrices into an overall similarity matrix for all participants.

\begin{figure}[t]
    \centering
    \includegraphics[trim = {0.5cm 0.4cm 0.03cm 0.1cm}, clip, width=0.95\linewidth]{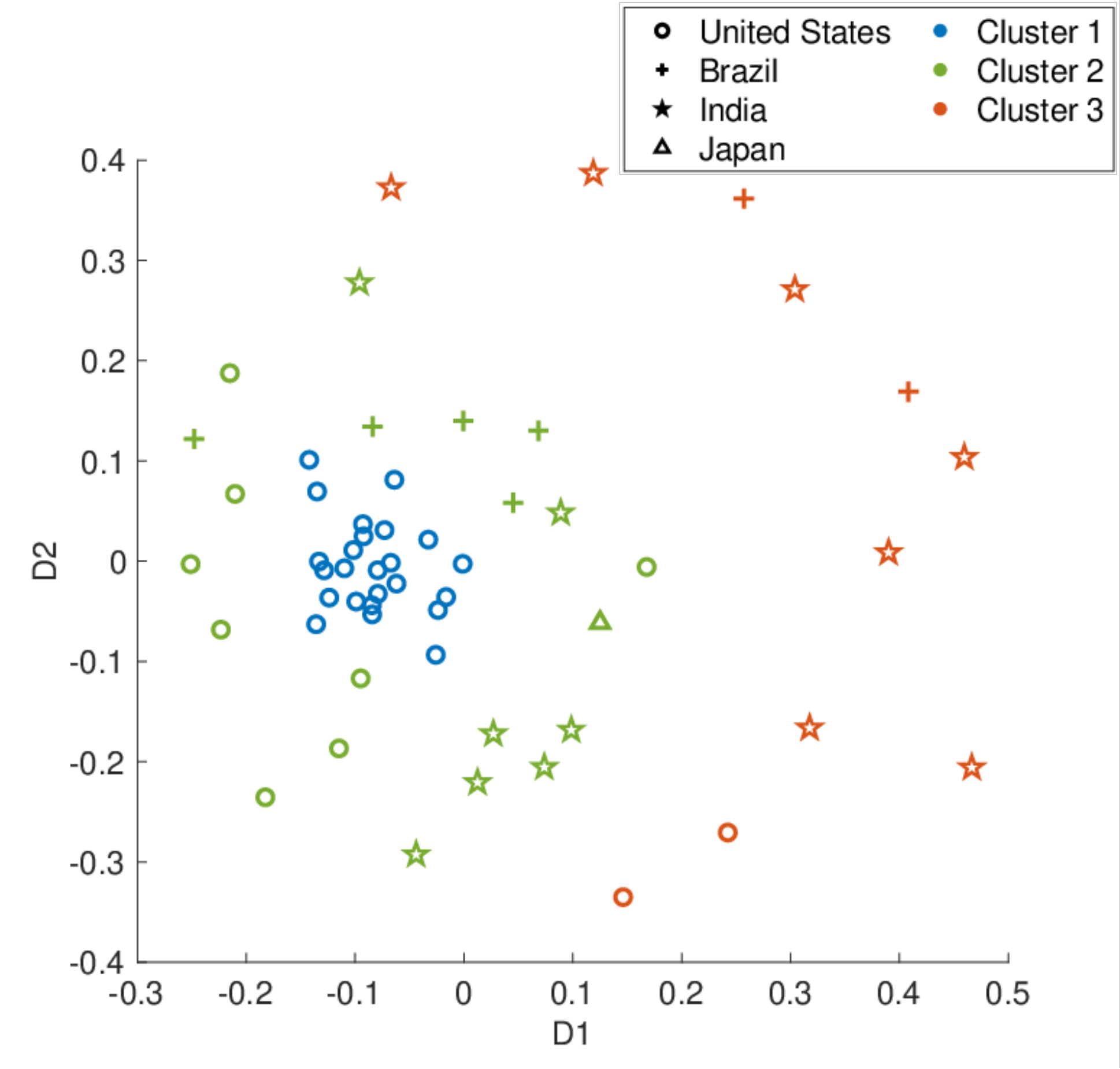}
    \caption{MDS plot visualizing similarity of the participants in their grouping of the gestures. Each mark represents one participant. The color and shape of the marks denote the results of clustering analysis and participant backgrounds, respectively. Participants in cluster 3 (red marks) were identified as potential outliers and were removed from further analysis.}
    \label{fig:outliermds}
\end{figure}

 We projected the participant similarities into two dimensions using non-metric Multidimensional Scaling (nMDS) and used clustering to assess outliers (Figure~\ref{fig:outliermds}). nMDS is a common dimensionality reduction technique for visualizing high-dimensional distances on a small number of dimensions. 
 Besides nMDS, we conducted k-means clustering with a range of 2 to 10 clusters on the dissimilarity matrix. The value of the Gap Statistic suggested 3 as the optimal number of clusters. Thus, we applied the k-means algorithm with 3 clusters to the data and visualized the clusters by using different colors for the participants in different clusters (Figure \ref{fig:outliermds}). 
Our analysis revealed that cluster 3 contained 11 out of the 16 participants that we had manually identified as potential outliers. Cluster 2 contained the remaining 5 potential outliers, as well as participants not considered to be outliers in the previous analysis. Thus, the two methods of manual and quantitative analysis of outliers largely overlapped and provided support that the cluster 3 participants either provided noisy data or judged similarities differently from the majority. Thus, we included the participants from clusters 1 and 2 ($n=45$ participants) in further analysis.

\begin{figure*}[tb]  
\centering
  \includegraphics[trim = {1cm, 8.4cm, 1.8cm, 9cm}, clip, width=0.85\textwidth]{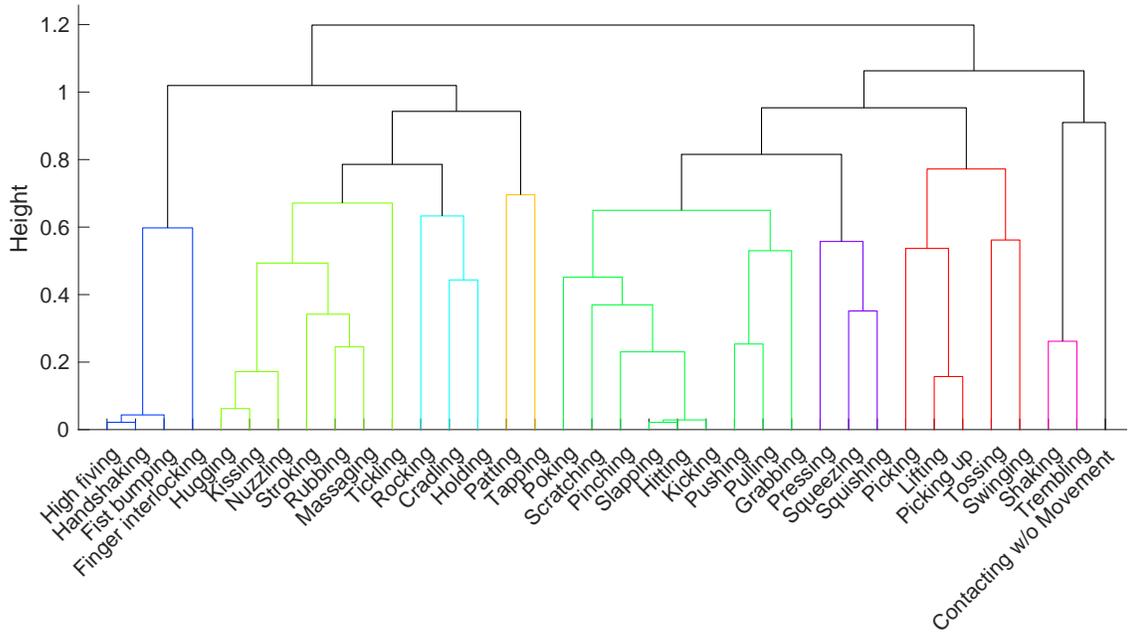}
  \caption{Results demonstrating hierarchical clustering results for the social touch gestures. The Gap Statistic criterion suggested an optimal number of 9 clusters. The Cophenetic correlation coefficient is 0.85 suggesting strong correspondence with the dissimilarity matrix. Each color represents one cluster.} 
  \label{fig:clusterplot}
\end{figure*}

The remaining participants were from the United States (32), followed by India (7), Brazil (5), and Japan (1). They self-identified as man ($n=29$), woman ($n=16$), or nonbinary ($n=0$). The mean age of the participants was 36.4 ($\pm 10.73$) years and their ages ranged between 21-63 years.  The participant background is denoted with the shape of the marks in Figure~\ref{fig:outliermds}. Participants who were not from the US are either in clusters 2 or 3. We analyzed this aspect further in our clustering results (Section~\ref{sec:clustering}).

\subsection{Coding descriptive names for the groups}
\label{sec:coding}
To understand the reasoning behind group choices, we coded the descriptive names provided by the participants for each group. From 4 to 8 to 12 groupings, the codes became more complex as subgroups began to form. The process of identifying these codes was iterative. For example, when coding the descriptive names for 12 groupings, we used the codes identified from 8 groupings in the first iteration. If we found any new or more specific patterns, we added new codes and recorded the previous data accordingly. Upon completing the coding of all the groupings, we had a total of 25 codes. We found some descriptive names to be ambiguous and coded them as `vague'. We also found that some names did not match the social touches they were assigned to, we coded these descriptive names as `random'. 
In some cases, participants labeled a group as `other' or `miscellaneous'. Thus, we also coded these groupings as `miscellaneous'. 
If a grouping contained only a single social touch, we coded it as `single action'. The remaining 21 codes included: aggressive, annoying, caregiving, direction, fingers, force, friendly, full-body, functional, grief, hands, holding onto, massage, nervous, playful, rapid, repetitive, romance, slow, social, and squeezing.

\subsection{Clustering touch gestures}
\label{sec:clustering}

Using the grouping data of each participant, we created a similarity matrix of touch gestures following the same procedure described by Russell~\cite{russel1980}. 
First, each pair of words was given a minimum similarity score of 1. If pairs of words were included in the same user-defined group, then their similarity score was increased by the number of groups being organized. For example, we increased the similarity score by 4 if a pair of words were in the same cluster for the 4 grouping mode for a participant. If a pair of words were included in the same group for 4, 8, and 12 groupings modes, then the words would have the maximum possible similarity of $1+4+8+12=25$. A single similarity matrix was calculated from the three grouping modes, and the matrix was subsequently normalized by dividing its entries by the maximum possible similarity value (i.e., 45 participants $\times$ 25 = 1125). We subtracted the normalized matrix from a matrix of ones to generate a dissimilarity matrix for all the gestures.

We applied clustering to the dissimilarity matrix and identified 9 clusters for the touch gestures.
Specifically, we employed agglomerative hierarchical clustering using the unweighted pair group method with arithmetic mean (UPGMA)
~\cite{murtagh2017algorithms}. To determine the optimal number of clusters for hierarchical clustering, we utilized the Gap Statistic evaluation criterion with a range of 2 to 10 clusters. This analysis suggested the presence of 9 clusters (Figure~\ref{fig:clusterplot}). 
The Cophenetic correlation coefficient was 0.85 for the 9 clusters, indicating a strong positive correspondence between the clusters and the original dissimilarity matrix. 
These clusters include: 
\vspace{0.2cm}
\begin{itemize}
\item \textbf{Cluster 1:} \textit{high-fiving}, \textit{handshaking}, \textit{fist bumping}, and \textit{finger interlocking}
\item \textbf{Cluster 2:} \textit{hugging}, \textit{kissing}, \textit{nuzzling}, \textit{stroking}, \textit{rubbing}, \textit{massaging}, and \textit{tickling} 
\item \textbf{Cluster 3:} \textit{rocking}, \textit{cradling}, and \textit{holding}
\item \textbf{Cluster 4:} \textit{patting} and \textit{tapping}
\item \textbf{Cluster 5:} \textit{poking}, \textit{scratching}, \textit{pinching}, \textit{slapping}, \textit{hitting}, \textit{kicking}, \textit{pushing}, \textit{pulling}, and \textit{grabbing}
\item \textbf{Cluster 6:} \textit{pressing}, \textit{squeezing}, and \textit{squishing}
\item \textbf{Cluster 7:} \textit{picking}, \textit{lifting up}, \textit{picking up}, \textit{tossing}, and \textit{swinging}
\item \textbf{Cluster 8:} \textit{shaking} and \textit{trembling}
\item \textbf{Cluster 9:} \textit{contacting without movement}
\end{itemize}
\vspace{0.2cm}

To test the effect of cultural background and English proficiency in our results, we repeated the above clustering analysis on data from 32 participants from the US. The analysis led to similar clusters with the exception that clusters 2 and 3 were merged into one cluster. Thus, we decided to continue with the above 9 clusters in our further analysis.

\begin{table}[t]
    \caption{Our derived names and the top 5 codes with their percentages for the 9 clusters. All the single-item groups are coded as `single action'. }
      \begin{tabularx}{0.486\textwidth}{*{5}{X}}
        \toprule
        \multicolumn{5}{l}{\textbf{Cluster 1: Social}} \vspace{0.1cm}\\ 
        \multicolumn{1}{c}{social} &
        \multicolumn{1}{c}{hands} &
        \multicolumn{1}{c}{romance} &
        \multicolumn{1}{c}{caregiving} &  
        \multicolumn{1}{c}{random} \\
        \multicolumn{1}{c}{50\%} &
        \multicolumn{1}{c}{14\%} &
        \multicolumn{1}{c}{14\%} &
        \multicolumn{1}{c}{4\%} &  
        \multicolumn{1}{c}{3\%} \\

        \midrule
        \multicolumn{5}{l}{\textbf{Cluster 2: Romantic Affection}} \vspace{0.1cm}\\ 
        \multicolumn{1}{c}{romance} &
        \multicolumn{1}{c}{caregiving} &
        \multicolumn{1}{c}{massage} &
        \multicolumn{1}{c}{random} &  
        \multicolumn{1}{c}{vague} \\
        \multicolumn{1}{c}{33\%} &
        \multicolumn{1}{c}{11\%} &
        \multicolumn{1}{c}{6\%} &
        \multicolumn{1}{c}{5\%} &  
        \multicolumn{1}{c}{4\%} \\

        \midrule

        \multicolumn{5}{l}{\textbf{Cluster 3: Caregiving Affection}} \vspace{0.1cm}\\ 
        \multicolumn{1}{c}{caregiving} &
        \multicolumn{1}{c}{romance} &
        \multicolumn{1}{c}{vague} &
        \multicolumn{1}{c}{functional} &  
        \multicolumn{1}{c}{random} \\
        \multicolumn{1}{c}{15\%} &
        \multicolumn{1}{c}{12\%} &
        \multicolumn{1}{c}{10\%} &
        \multicolumn{1}{c}{8\%} &  
        \multicolumn{1}{c}{6\%} \\

        \midrule

        \multicolumn{5}{l}{\textbf{Cluster 4: Hand Contact}} \vspace{0.1cm}\\ 
        \multicolumn{1}{c}{force} &
        \multicolumn{1}{c}{hands} &
        \multicolumn{1}{c}{social} &
        \multicolumn{1}{c}{vague} &  
        \multicolumn{1}{c}{random} \\
        \multicolumn{1}{c}{12\%} &
        \multicolumn{1}{c}{10\%} &
        \multicolumn{1}{c}{9\%} &
        \multicolumn{1}{c}{9\%} &  
        \multicolumn{1}{c}{8\%} \\ 

         \midrule

        \multicolumn{5}{l}{\textbf{Cluster 5: Aggression}} \vspace{0.1cm}\\ 
        \multicolumn{1}{c}{aggressive} &
        \multicolumn{1}{c}{functional} &
        \multicolumn{1}{c}{random} &
        \multicolumn{1}{c}{hands} &  
        \multicolumn{1}{c}{vague} \\
        \multicolumn{1}{c}{52\%} &
        \multicolumn{1}{c}{9\%} &
        \multicolumn{1}{c}{5\%} &
        \multicolumn{1}{c}{5\%} &  
        \multicolumn{1}{c}{5\%} \\

         \midrule

        \multicolumn{5}{l}{\textbf{Cluster 6: Forceful Press}} \vspace{0.1cm} \\ 
        \multicolumn{1}{c}{aggressive} &
        \multicolumn{1}{c}{functional} &
        \multicolumn{1}{c}{squeezing} &
        \multicolumn{1}{c}{vague} &  
        \multicolumn{1}{c}{force} \\
        \multicolumn{1}{c}{18\%} &
        \multicolumn{1}{c}{12\%} &
        \multicolumn{1}{c}{12\%} &
        \multicolumn{1}{c}{9\%} &  
        \multicolumn{1}{c}{8\%} \\ 

        \midrule
        
        \multicolumn{5}{l}{\textbf{Cluster 7: Functional Movement}} \vspace{0.1cm}\\ 
        \multicolumn{1}{c}{functional} &
        \multicolumn{1}{c}{vague} &
        \multicolumn{1}{c}{aggressive} &
        \multicolumn{1}{c}{random} &  
        \multicolumn{1}{c}{hands} \\
        \multicolumn{1}{c}{29\%} &
        \multicolumn{1}{c}{12\%} &
        \multicolumn{1}{c}{8\%} &
        \multicolumn{1}{c}{8\%} &  
        \multicolumn{1}{c}{5\%} \\ 

        \midrule
        
        \multicolumn{5}{l}{\textbf{Cluster 8: Nervous Contact}} \vspace{0.1cm}\\
        \multicolumn{1}{c}{nervous} &
        \multicolumn{1}{c}{aggressive} &
        \multicolumn{1}{c}{vague} &
        \multicolumn{1}{c}{random} &  
        \multicolumn{1}{c}{force} \\
        \multicolumn{1}{c}{30\%} &
        \multicolumn{1}{c}{14\%} &
        \multicolumn{1}{c}{11\%} &
        \multicolumn{1}{c}{6\%} &  
        \multicolumn{1}{c}{5\%} \\ 

        \midrule
        
        \multicolumn{5}{l}{\textbf{Cluster 9: Contact w/o Movement}} \vspace{0.1cm}\\ 
        \multicolumn{1}{c}{single action} &
        \multicolumn{1}{c}{miscellaneous} &
        \multicolumn{1}{c}{social} &
        \multicolumn{1}{c}{vague} &  
        \multicolumn{1}{c}{functional} \\
        \multicolumn{1}{c}{24\%} &
        \multicolumn{1}{c}{10\%} &
        \multicolumn{1}{c}{10\%} &
        \multicolumn{1}{c}{7\%} &  
        \multicolumn{1}{c}{6\%} \\ 
       
        \bottomrule
    \end{tabularx}
    \label{tab:code-freq}
\end{table}

\subsection{Interpreting the clusters}
We calculated the distribution of our codes for the descriptive names across these clusters to interpret the reason behind the groups. Table~\ref{tab:code-freq} presents the five frequent codes for the gestures in each cluster. 

We named the clusters based on the distribution of their five top codes.
For clusters 1 and 5, the majority of the codes ($\ge50\%$) are `social' and `aggressive'. Thus, we call these clusters \textit{Social} and \textit{Aggression} respectively. 
Clusters 2, 7, 8, and 9 have one frequent code ($\ge24\%$), followed by one or two  codes with $\ge10\%$ frequency. 
For cluster 2, the top code is `romance' followed by `caregiving', both of which reflect the affective nature of touch. Thus, we name this group as \textit{Romantic Affection}. For cluster 7, the top code is `functional', followed by `vague'. This cluster includes a set of gestures that involve lifting and moving an object or person. Thus, we name it \textit{Functional Movement}. 
Cluster 8 has a top code of `nervous', followed by `aggressive'. Thus, we call it \textit{Nervous Contact}. Cluster 9 includes the single gesture of \textit{contacting without movement}. This gesture was often put in a separate group by the participants and we coded it as `single action'. Thus, we name this cluster as \textit{Contact w/o Movement} to reflect its distinct nature in the participants' minds. 
Finally, clusters 3, 4, and 6 have a relatively flat code distribution. Cluster 3 has the same two top codes as cluster 2, representing affect, but in the reverse order. Thus, we name it \textit{Caregiving Affection}. 
Cluster 4 has two codes of `force' and `hands' with more than 10\% frequency. With two gestures of \textit{patting} and \textit{tapping}, we name this cluster as \textit{Hand Contact}. The top codes ($\ge10\%$) for cluster 6 are `aggressive', `functional', and `squeezing'. Since the top labels indicate both the `aggressive' and `functional' aspects of the gestures in this cluster, we use a neutral label and call this cluster \textit{Forceful Press}. 
  
We discuss these clusters and their implications for HRI research in the next section.

\section{Discussion}

In this study, we present data on the user perception and description of touch gestures. Our findings indicate that users tend to assess the similarity of touch gestures based on their emotional and social connotations, in addition to the functional and contact characteristics. Specifically, cluster 1 includes touch gestures that are frequently annotated with `social' names. Clusters 2 and 3 include gestures that are mainly coded with positive associations of `romance' and `caregiving'. Similarly, clusters 5 and 8 are coded with negative descriptors of `aggressive' and `nervous'. Finally, four clusters (i.e., 4, 6, 7, 9) seem to be mainly described based on the characteristics of the contact such as the body part (cluster 4), force (cluster 6), and whether the touch involved movement (cluster 7) or not (cluster 9). 
These clusters emerged without providing information on the context of interaction, suggesting that users have strong social, positive, negative, and functional associations with touch gestures even without context. 
Some clusters have a flat distribution of codes and show a notable mix of affective and functional interpretations (e.g., cluster 6 with \textit{pressing}, \textit{squeezing}, and \textit{squishing}) suggesting that an individual's background or interaction context may notably shift their meaning.
Interestingly \textit{contacting without movement} was often regarded as different from the other gestures, which could be due to its neutral emotional content as well as the static nature of the touch.

 These user-generated clusters are a step toward a framework for the analysis and understanding of social touch and can inform research on sensing, designing, and analyzing human-robot touch interactions. We anticipate the following use cases of the touch clusters for HRI:

\vspace{0.2cm}
\noindent (1) \textit{Sensing touch from humans.} 
A desirable factor for robotic touch sensors is their ability to recognize a variety of gestures~\cite{jung2017automatic}. These clusters can aid researchers in selecting gestures that are different in their semantic and contact characteristics. For instance, the co-location of \textit{stroking} and \textit{rubbing} gestures in cluster 2 suggests that it might be appropriate to choose one of the two gestures.
Relatedly, when evaluating the efficacy of a touch-sensing algorithm~\cite{jung2017automatic,burns2022touchperception,Choi2022Learning}, HRI researchers can weigh misclassifications according to these semantic clusters. For example, misclassifying \textit{stroking} with \textit{slapping} should be penalized more than mistaking \textit{stroking} with \textit{rubbing} or \textit{nuzzling}. 

\vspace{0.2cm}
\noindent (2) \textit{Interpreting and responding to touch from humans.} The proposed touch gesture clusters can aid robots in responding intelligently to human touch. These clusters can help robots identify the intention behind touch gestures.
While the significance and purpose of social touch gestures may depend on the context, these clusters and their labels can help develop a probabilistic mental model for robots about a user's intent of a touch gesture. During an interaction episode, the robot can update these probabilities based on other contextual parameters such as the user's verbal utterances. 

\vspace{0.2cm}
\noindent (3) \textit{Touching people to communicate.} The semantic clusters can help design and evaluate robots that touch humans to communicate information or emotion~\cite{teyssier2020conveying}. Specifically, to evaluate the efficacy of a robot in applying touch gestures to humans,
HRI researchers can determine the degree of dissimilarity between the intended touch gesture and the one identified by the human.
Also, depending on the purpose of the interaction (e.g., social, emotional, or functional), the robot may use the clusters to select and use alternative gestures with similar connotations.

\vspace{0.2cm}
\noindent (4) \textit{Analyzing human-robot touch interactions.} HRI researchers can use these clusters to code video recordings of touch interactions with a robot and aggregate touch interaction into higher-level themes. To support this, our work builds on the touch dictionary~\cite{yohanan2012role} by providing data on the relationship between touch gestures. Thus, these clusters provide an initial framework for the analysis of social touch interactions with robots.

\section{Conclusion and Future Work}

Our work is a first step toward charting the relationship of touch gestures for HRI. We anticipate that our results can pave the way for future work on designing and evaluating robots that use touch as a non-verbal communication channel.

We see several avenues for extending this work. First, the relationship between the user-generated clusters for touch gestures and signals produced by the gestures on different touch sensors is an open question. A good touch sensor should be able to create distinct signals for gestures that are in different clusters according to user perception. Also, robots should be able to create distinct sensations when touching users with gestures in different clusters. d

Second, future work can examine the impact of presentation modality on the semantic relationship of touch gestures. In this paper, we presented text labels for social touch gestures, following the common procedure in user studies of touch sensing for social robots. This approach helped abstract different styles of applying the gestures and study the user's mental representations of archetypal touch gestures.
Future studies can examine how people group the touch gestures using videos or by applying robot touch on the user's body and compare the results to the clusters we found in this work. These studies should capture a wide range of touch styles (e.g., contact, force) for each gesture to avoid biasing the results to a small sample.

Finally, the meaning of touch can vary based on contexts, cultures, and individuals. As a first step, we examined if any generalizable patterns could be found about the relationships between various touch gestures. Our study population primarily consisted of individuals that grew up in the United States. Participants from other cultures often fall into cluster 3 and around the borders of cluster 2. It is unclear whether this result is due to their familiarity with touch labels or the difference in their cultural background. Future studies can examine how the clusters of social touch gestures differ across cultures by translating the text labels into different languages. A larger dataset can also allow future work to look into individual differences in perception of social touch.

\bibliographystyle{ieeetr}
\bibliography{references}

\begin{thebibliography}{10}

\bibitem{huisman2017social}
G.~Huisman, ``Social touch technology: A survey of haptic technology for social
  touch,'' {\em IEEE Transactions on Haptics}, vol.~10, no.~3, pp.~391--408,
  2017.

\bibitem{yohanan2012role}
S.~Yohanan and K.~E. MacLean, ``The role of affective touch in human-robot
  interaction: Human intent and expectations in touching the haptic creature,''
  {\em International Journal of Social Robotics}, vol.~4, no.~2, pp.~163--180,
  2012.

\bibitem{fitter2014analyzing}
N.~T. Fitter and K.~J. Kuchenbecker, ``Analyzing human high-fives to create an
  effective high-fiving robot,'' in {\em Proceedings of the ACM/IEEE
  International Conference on Human-Robot Interaction (HRI)}, pp.~156--157,
  2014.

\bibitem{burns2022touchperception}
R.~B. Burns, H.~Lee, H.~Seifi, R.~Faulkner, and K.~J. Kuchenbecker, ``Endowing
  a nao robot with practical social-touch perception,'' {\em Frontiers in
  Robotics and AI}, p.~86, 2022.

\bibitem{hertenstein2009}
M.~J. Hertenstein, R.~Holmes, M.~McCullough, and D.~Keltner, ``The
  communication of emotion via touch.,'' {\em Emotion}, vol.~9, no.~4, p.~566,
  2009.

\bibitem{jung2014touching}
M.~M. Jung, R.~Poppe, M.~Poel, and D.~K. Heylen, ``Touching the
  void--introducing cost: corpus of social touch,'' in {\em Proceedings of the
  International Conference on Multimodal Interaction (ICMI)}, pp.~120--127,
  2014.

\bibitem{russel1980}
J.~Russell, ``A circumplex model of affect,'' {\em Journal of Personality and
  Social Psychology}, vol.~39, pp.~1161--1178, 12 1980.

\bibitem{xu2021subtle}
S.~Xu, C.~Xu, S.~McIntyre, H.~Olausson, and G.~J. Gerling, ``Subtle contact
  nuances in the delivery of human-to-human touch distinguish emotional
  sentiment,'' {\em IEEE Transactions on Haptics}, vol.~15, no.~1, pp.~97--102,
  2021.

\bibitem{cang2015different}
X.~L. Cang, P.~Bucci, A.~Strang, J.~Allen, K.~MacLean, and H.~S. Liu,
  ``Different strokes and different folks: Economical dynamic surface sensing
  and affect-related touch recognition,'' in {\em Proceedings of the ACM
  International Conference on Multimodal Interaction (ICMI)}, pp.~147--154,
  2015.

\bibitem{andreasson2018affective}
R.~Andreasson, B.~Alenljung, E.~Billing, and R.~Lowe, ``Affective touch in
  human--robot interaction: conveying emotion to the nao robot,'' {\em
  International Journal of Social Robotics}, vol.~10, pp.~473--491, 2018.

\bibitem{willemse2019social}
C.~J. Willemse and J.~B. Van~Erp, ``Social touch in human--robot interaction:
  Robot-initiated touches can induce positive responses without extensive prior
  bonding,'' {\em International Journal of Social Robotics}, vol.~11, no.~2,
  pp.~285--304, 2019.

\bibitem{teyssier2020conveying}
M.~Teyssier, G.~Bailly, C.~Pelachaud, and E.~Lecolinet, ``Conveying emotions
  through device-initiated touch,'' {\em IEEE Transactions on Affective
  Computing}, 2020.

\bibitem{seifi2023firsthand}
H.~Seifi, S.~A. Vasquez, H.~Kim, and P.~Fazli, ``First-hand impressions:
  Charting and predicting user impressions of robot hands,'' {\em ACM
  Transactions on Human-Robot Interaction}, 2023.

\bibitem{Choi2022Learning}
H.~Choi, D.~Brouwer, M.~A. Lin, K.~T. Yoshida, C.~Rognon, B.~Stephens-Fripp,
  A.~M. Okamura, and M.~R. Cutkosky, ``Deep learning classification of touch
  gestures using distributed normal and shear force,'' in {\em Proceedings of
  the IEEE/RSJ International Conference on Intelligent Robots and Systems
  (IROS)}, pp.~3659--3665, 2022.

\bibitem{teyssier2021human}
M.~Teyssier, B.~Parilusyan, A.~Roudaut, and J.~Steimle, ``Human-like artificial
  skin sensor for physical human-robot interaction,'' in {\em Proceedings of
  the IEEE International Conference on Robotics and Automation (ICRA)},
  pp.~3626--3633, IEEE, 2021.

\bibitem{jung2017automatic}
M.~M. Jung, M.~Poel, R.~Poppe, and D.~K. Heylen, ``Automatic recognition of
  touch gestures in the corpus of social touch,'' {\em Journal on Multimodal
  User Interfaces}, vol.~11, no.~1, pp.~81--96, 2017.

\bibitem{pelikan2020you}
H.~R. Pelikan, M.~Broth, and L.~Keevallik, ``" are you sad, cozmo?" how humans
  make sense of a home robot's emotion displays,'' in {\em Proceedings of the
  ACM/IEEE International Conference on Human-Robot Interaction (HRI)},
  pp.~461--470, 2020.

\bibitem{rognon2022online}
C.~Rognon, T.~Bunge, M.~Gao, C.~Conor, B.~Stephens-Fripp, C.~Brown, and
  A.~Israr, ``An online survey on the perception of mediated social touch
  interaction and device design,'' {\em IEEE Transactions on Haptics}, vol.~15,
  no.~2, pp.~372--381, 2022.

\bibitem{prasad2022mild}
V.~Prasad, D.~Koert, R.~Stock-Homburg, J.~Peters, and G.~Chalvatzaki, ``Mild:
  Multimodal interactive latent dynamics for learning human-robot
  interaction,'' in {\em Proceedings of the IEEE-RAS International Conference
  on Humanoid Robots (Humanoids)}, pp.~472--479, 2022.

\bibitem{wang2011handshake}
Z.~Wang, E.~Giannopoulos, M.~Slater, and A.~Peer, ``Handshake: Realistic
  human-robot interaction in haptic enhanced virtual reality,'' {\em Presence},
  vol.~20, no.~4, pp.~371--392, 2011.

\bibitem{ammi2015haptic}
M.~Ammi, V.~Demulier, S.~Caillou, Y.~Gaffary, Y.~Tsalamlal, J.-C. Martin, and
  A.~Tapus, ``Haptic human-robot affective interaction in a handshaking social
  protocol,'' in {\em Proceedings of the ACM/IEEE International Conference on
  Human-Robot Interaction (HRI)}, pp.~263--270, 2015.

\bibitem{salter2007using}
T.~Salter, F.~Michaud, D.~L{\'e}tourneau, D.~Lee, and I.~P. Werry, ``Using
  proprioceptive sensors for categorizing human-robot interactions,'' in {\em
  Proceedings of the ACM/IEEE International Conference on Human-Robot
  Interaction (HRI)}, pp.~105--112, 2007.

\bibitem{li2016human}
B.~Li, L.~Boccanfuso, Q.~Wang, E.~Barney, Y.~A. Ahn, C.~Foster, K.~Chawarska,
  B.~Scassellati, and F.~Shic, ``Human robot activity classification based on
  accelerometer and gyroscope,'' in {\em Proceedings of the IEEE International
  Symposium on Robot and Human Interactive Communication (RO-MAN)},
  pp.~423--424, 2016.

\bibitem{eurohaptics22}
R.~Abou~Chahine, D.~Kwon, C.~Lim, G.~Park, and H.~Seifi, ``Vibrotactile
  similarity perception in crowdsourced and lab studies,'' in {\em Proceedings
  of the International Conference on Human Haptic Sensing and Touch Enabled
  Computer Applications (EuroHaptics)}, pp.~255--263, Springer, 2022.

\bibitem{park2011perceptual}
G.~Park and S.~Choi, ``Perceptual space of amplitude-modulated vibrotactile
  stimuli,'' in {\em Proceedings of the IEEE World Haptics Conference (WHC)},
  pp.~59--64, IEEE, 2011.

\bibitem{tsogo2000multidimensional}
L.~Tsogo, M.~Masson, and A.~Bardot, ``Multidimensional scaling methods for
  many-object sets: A review,'' {\em Multivariate Behavioral Research},
  vol.~35, no.~3, pp.~307--319, 2000.

\bibitem{spam2013versatility}
M.~C. Hout, S.~D. Goldinger, and R.~W. Ferguson, ``The versatility of spam: A
  fast, efficient, spatial method of data collection for multidimensional
  scaling.,'' {\em Journal of Experimental Psychology: General}, vol.~142,
  no.~1, p.~256, 2013.

\bibitem{ternes2008designing}
D.~Ternes and K.~E. MacLean, ``Designing large sets of haptic icons with
  rhythm,'' in {\em Proceedings of the International Conference on Human Haptic
  Sensing and Touch Enabled Computer Applications (EuroHaptics)}, pp.~199--208,
  Springer, 2008.

\bibitem{rugg1997sorting}
G.~Rugg and P.~McGeorge, ``The sorting techniques: a tutorial paper on card
  sorts, picture sorts and item sorts,'' {\em Expert Systems}, vol.~14, no.~2,
  pp.~80--93, 1997.

\bibitem{seifi2020capturing}
H.~Seifi, M.~Oppermann, J.~Bullard, K.~E. MacLean, and K.~J. Kuchenbecker,
  ``Capturing experts' mental models to organize a collection of haptic
  devices: Affordances outweigh attributes,'' in {\em Proceedings of the ACM
  SIGCHI Conference on Human Factors in Computing Systems (CHI)}, pp.~1--12,
  2020.

\bibitem{mun2019perceptual}
S.~Mun, H.~Lee, and S.~Choi, ``Perceptual space of regular homogeneous haptic
  textures rendered using electrovibration,'' in {\em Proceedings of the IEEE
  World Haptics Conference (WHC)}, pp.~7--12, 2019.

\bibitem{silvera2012interpretation}
D.~Silvera~Tawil, D.~Rye, and M.~Velonaki, ``Interpretation of the modality of
  touch on an artificial arm covered with an eit-based sensitive skin,'' {\em
  The International Journal of Robotics Research}, vol.~31, no.~13,
  pp.~1627--1641, 2012.

\bibitem{hertenstein2011gender}
M.~J. Hertenstein and D.~Keltner, ``Gender and the communication of emotion via
  touch,'' {\em Sex roles}, vol.~64, pp.~70--80, 2011.

\bibitem{cramer2009effects}
H.~S. Cramer, N.~A. Kemper, A.~Amin, and V.~Evers, ``The effects of robot touch
  and proactive behaviour on perceptions of human-robot interactions.,'' in
  {\em Proceedings of the ACM/IEEE International Conference on Human-Robot
  Interaction (HRI)}, pp.~275--276, 2009.

\bibitem{burns2021getting}
R.~B. Burns, H.~Seifi, H.~Lee, and K.~J. Kuchenbecker, ``Getting in touch with
  children with autism: Specialist guidelines for a touch-perceiving robot,''
  {\em Paladyn, Journal of Behavioral Robotics}, vol.~12, no.~1, pp.~115--135,
  2021.

\bibitem{salter2006learning}
T.~Salter, K.~Dautenhahn, and R.~te~Boekhorst, ``Learning about natural
  human--robot interaction styles,'' {\em Robotics and Autonomous Systems},
  vol.~54, no.~2, pp.~127--134, 2006.

\bibitem{murtagh2017algorithms}
F.~Murtagh and P.~Contreras, ``Algorithms for hierarchical clustering: an
  overview, ii,'' {\em Wiley Interdisciplinary Reviews: Data Mining and
  Knowledge Discovery}, vol.~7, no.~6, p.~e1219, 2017.

\end{thebibliography}

\end{document}